\documentclass[sigconf]{acmart}

\AtBeginDocument{%
  }

\setcopyright{acmlicensed}
\copyrightyear{2027}
\acmYear{2027}
\acmDOI{XXXXXXX.XXXXXXX}
\acmConference[KDD '27]{The 33rd ACM SIGKDD Conference on Knowledge Discovery and Data Mining}{August 2027}{San Jose, CA, USA}

\usepackage[british]{babel}

\usepackage{amsmath}
\usepackage{amsfonts}
\usepackage{mathrsfs}

\usepackage{algorithm}
\usepackage[noend]{algpseudocode}  

\usepackage{graphicx}
\usepackage{subcaption}

\usepackage{booktabs}      
\usepackage{threeparttable}
\usepackage{diagbox}
\usepackage{listings}
\usepackage{enumitem}
\usepackage{csquotes}      
\usepackage{comment}
\usepackage{url}

\begin{document}

\title{Calibrate Globally, Measure Everywhere: Scaling LLM-Based Prevalence Measurement Across A/B Experiments}

\author{Zehao Xu}
\affiliation{%
  \institution{Pinterest, Inc.}
  \city{Toronto}
  \state{ON}
  \country{Canada}
}
\email{zehaoxu@pinterest.com}

\author{Tony Paek}
\affiliation{%
  \institution{Pinterest, Inc.}
  \city{New York}
  \state{NY}
  \country{USA}
}
\email{tpaek@pinterest.com}

\author{Kevin O'Sullivan}
\affiliation{%
  \institution{Pinterest, Inc.}
  \city{New York}
  \state{NY}
  \country{USA}
}
\email{kosullivan@pinterest.com}

\author{Attila Dobi}
\affiliation{%
  \institution{Pinterest, Inc.}
  \city{San Francisco}
  \state{CA}
  \country{USA}
}
\email{adobi@pinterest.com}
\begin{abstract}
Online media platforms often track the share of impressions associated with specific content attributes, or prevalence, to evaluate trade-offs and set guardrails in A/B experiments. LLM-based labeling provides a high-fidelity reference measurement, but is cost-prohibitive to run per experiment, per arm, per segment, and per day on a platform with hundreds of concurrent experiments. 

We describe a surrogate-based prevalence measurement system deployed in Pinterest's experimentation platform. The contribution is system-level rather than estimator-level: the system maintains a single global calibration of ML score buckets, continuously refreshed from a recurring LLM-labeled stream, and reuses the resulting bucket-level prevalences across every experiment via a per-experiment SQL metric and a delta-focused dashboard. Because the calibration is derived from the platform's daily-batch prevalence samples, it remains representative of production traffic as distributions drift, and in Pinterest's deployment it incurs zero incremental labeling cost. Teams without such infrastructure can instantiate the same pattern with a recurring calibration-labeling workflow whose cost is amortized across all downstream experiments rather than paid per experiment, arm, segment, and day. 

The system serves $\sim$100 experiments and $\sim$250 arms per day across six calibrated content categories, including the company-wide holdout program. 
Relative to per-experiment LLM labeling, which in practice yields a single one-shot read per arm on a small subset of experiments, the surrogate provides daily per-arm prevalence on over 20$\times$ as many concurrent arms under the same labeling budget. Across roughly 300 production audits, the surrogate's 95\% confidence interval contains the LLM-based reference point estimate in 92\% of evaluations, and day-level delta aggregation recovers small 2--5\% relative shifts that no single per-arm LLM measurement can detect.
\end{abstract}





\maketitle

\section{Introduction}

Modern recommender systems and media platforms must balance user
engagement against the need to manage exposure to certain content
attributes.
Teams often summarize such exposure objectives in terms of
\emph{prevalence}: the fraction of impressions
associated with a given target category. At the same time, product
decisions are largely driven by large scale A/B
experiments, where variants adjust ranking or
filtering and are evaluated on both engagement and attribute specific
exposure metrics.

One way to measure prevalence is to sample content from
traffic, label it with LLMs using expert validated prompts, and
apply a design consistent estimator. In our setting, we use PPSWOR
(probability proportional to size without replacement) sampling~\cite{horvitz1952generalization} and the
Hansen--Hurwitz estimator~\cite{hansen1943theory} to obtain
high quality and unbiased measurements.

However, directly using LLM-based prevalence as a default metric for
every experiment is impractical. Running a separate LLM job per arm and
per segment is expensive, and quickly becomes infeasible on a
platform with hundreds of concurrent experiments. Moreover,
experimenters often care about relatively small but meaningful changes
in prevalence. A single LLM measurement per arm tends to emphasize the
\emph{absolute} level, so many small
treatment-control deltas appear statistically non-significant. Running LLM labeling
per experiment, per segment, and per day to address this would further
amplify cost.

The contribution of this paper is not a new prevalence estimator but a deployed measurement system. We adopt an ML-score surrogate that turns a recurring global LLM-labeled calibration sample into reusable bucket-level prevalences: model scores are discretized into buckets, bucket-level prevalences are calibrated once, and
per-experiment estimates are then computed from impression logs alone via a deterministic SQL metric and a delta-focused dashboard. In Pinterest's deployment, this calibration sample is supplied by an existing daily LLM-based prevalence workflow for platform-level measurement, so the surrogate incurs zero incremental labeling cost.

The system has been deployed in Pinterest's experimentation platform since January 2026, currently serving
$\sim$100 experiments and $\sim$250 arms per day across the
Content Quality, Trust \& Safety, and Overall Holdout programs.
We state the methodology in Section~\ref{sec:methodology}, audit against the
LLM-based reference on live A/B experiments in Section~\ref{sec:validation}, describe system implementation in Section~\ref{sec:implementation}, and discuss design
choices and the LLM cost outlook
in Section~\ref{sec:discussion}.

\section{Related Work}
\label{sec:related_work}

Our work sits at the intersection of five strands of prior work,
and combines pieces from each into a deployed measurement system
for online experimentation.

\textbf{Surrogate outcomes.} Surrogate outcome methods replace an
expensive or slow-to-observe outcome with a cheaper proxy that is
correlated with it~\cite{prentice1989surrogate}. Recent work has
extended this idea to treatment effect estimation in the
short run/long run setting~\cite{athey2025surrogate}. We adopt
this framing operationally: a fast model score serves as the
surrogate, and an expensive LLM labeled measurement serves as the
reference target the surrogate is calibrated against and
periodically audited by.

\textbf{Post-stratification and calibrated estimation.} The
underlying estimator is a post-stratification of model scores
into discrete buckets, calibrated on an LLM labeled
sample~\cite{holt1979post,sarndal1992model}. Compared with direct
Hansen--Hurwitz estimation~\cite{hansen1943theory} on a
per experiment sample, post stratification reuses the calibration
across experiments and segments, which is what makes the cost
amortization in this paper possible.

\textbf{Online experimentation and variance reduction.}
Large scale A/B testing has produced a rich literature on
variance reduction for treatment effect estimation, of which
CUPED~\cite{deng2013improving} is the most widely deployed
example. CUPED uses pre experiment covariates to reduce variance
in the metric of interest. Our method addresses a different but
complementary problem: the metric itself is too costly to
\emph{compute} per experiment, not just too noisy. The two ideas
are composable; we leave a CUPED on $\Delta(d)$ extension to
future work.

\textbf{Score calibration.} Mapping classifier scores to
calibrated probabilities is a long studied
problem~\cite{platt1999probabilistic, zadrozny2002transforming}.
Bucketed ("histogram") calibration is the simplest such mapping
and is convenient for our setting because it is computed once
offline and consumed via a simple table lookup at SQL query time.
We treat the choice of bucket scheme as a deployment decision
rather than a contribution; alternatives are discussed in
Section~\ref{sec:bucket_choice}.

\textbf{LLM-based labeling.} Prompted large language models,
      optionally with expert reviewed prompts and calibration are increasingly used as scalable judges in practice~\cite{openai2023gpt4moderation,zheng2023mtbench,liu2023_mmsafetybench}. At Pinterest, an LLM labeling pipeline of this form is in production as the reference quality measurement system for content attribute prevalence~\cite{kdd_paper_2026}.

\section{Prevalence Estimation}
\label{sec:iw_prevalence}

We consider the problem of estimating the prevalence of a category $k$ on a large scale media platform. Let
$$
  \mathcal{K} = \{\text{Food-Recipe}, \text{Lawn and Garden}, \text{Gen-AI Generated}, ...\}
$$
denote the set of content attributes considered in this study, and
fix a particular $k \in \mathcal{K}$.

Our target quantity is the prevalence of
category $k$ within $S$,  which can denote the full population or a
specific subset such as an experiment arm (control, treatment),
a user demographic (country, age), an app surface, or intersections of these. 

The statistical design and engineering details of our production
prevalence pipeline are described in Dobi et al.~\cite{kdd_paper_2026}. In this
section, we briefly recap the core estimator and introduce notation
that we will use throughout the rest of the paper, in particular when
we describe the deployed ML score surrogate method and its calibration.

\subsection{Notation}
We consider a large population of content items,
indexed by $i = 1,\dots,N$. For each item $i$, we observe:
\begin{itemize}
  \item $I_i$: the total number of impressions of item $i$ over a given
        time window.
  \item $Z_{i,k} \in \{0,1\}$: a label indicating whether item $i$ is truly in category $k$ ($1$) or not ($0$).
\end{itemize}

For a given segment $S$, let $\mathcal{D}(S) \subseteq \{1,\dots,N\}$
denote the set of items that receive impressions from $S$, and let
$I_i(S)$ be the number of impressions of item $i$ from $S$ over the
time window of interest. The total impressions of item $i$ satisfy
$I_i = \sum_{S} I_i(S)$. The prevalence of category $k$ in segment
$S$ can be written as:

\begin{equation}
\label{eq:prevalence_def}
\begin{aligned}
   \mathcal{P}_k(S)
    &=   \mathbb{P}\!\left(Z_{k} = 1 \mid \text{impressions} \in S\right) \\
    &=  \frac{\sum_{i \in \mathcal{D}(S)} Z_{i,k} I_i(S)}
            {\sum_{i \in \mathcal{D}(S)} I_i(S)},
\end{aligned}
\end{equation}
When $S$ is the entire population, $\mathcal{D}(S) = \{1,\dots,N\}$ and
$I_i(S) = I_i$, recovering the global prevalence definition.

\subsection{Sampling Design and Hansen--Hurwitz Estimator}
\label{sec:hh_estimator}

Directly labeling \emph{all} items in $\mathcal{D}(S)$ (often billions
of items across many segments) is infeasible, so we estimate these
quantities from a sample. We adopt a PPSWOR 
design, in which items are sampled with probabilities proportional to a
chosen size measure. Compared with simple random sampling, PPSWOR allows
us to oversample items that are more informative (e.g., high impression items), achieving similar statistical power with a much
smaller sample size.

For category $k$ and segment $S$, each item $i \in \mathcal{D}(S)$ is
assigned a sampling weight $w_{i,k}(S) \propto f\bigl(I_i(S)\bigr)$, and is selected into the sample with probability
$p_{sample,i,k}(S) = w_{i,k}(S) / \sum_{j \in \mathcal{D}(S)} w_{j,k}(S)$.
Here $f$ is a function of impressions; it can be $I_i(S)$ itself, or a
combination, such as $I_i(S)$ multiplied by other factors (e.g.,  model score).
We discuss concrete choices of $f$ in more detail in
Section~\ref{sec:offline_calibration}.

Suppose we draw a sample of size $n$ from $\mathcal{D}(S)$, with sampled indices $\{i_1, \ldots, i_n\}$. The Hansen--Hurwitz estimator for
the total target category impressions in $S$ and the prevalence in segment $S$ are estimated as
\begin{equation}
  \widehat{T}_k(S)
    \;=\;
    \frac{1}{n}
    \sum_{t=1}^n
      \frac{Z_{i_t,k} I_{i_t}(S)}{p_{\text{sample}, i_t,k}(S)}, \ \ \widehat{\mathcal{P}}_k(S)
    \;=\;
    \frac{\widehat{T}_k(S)}
         {\sum_{j \in \mathcal{D}(S)} I_j(S)}.
  \label{eq:prev_estimator_segment}
\end{equation}

In practice, both $\sum_{j \in \mathcal{D}(S)} I_j(S)$ 
and $\sum_{j \in \mathcal{D}(S)} w_{j,k}(S)$
are computed directly from
the underlying logs for segment $S$.
Strictly, the Hansen--Hurwitz form is the with replacement estimator; under PPSWOR with small per item inclusion probabilities it serves as the standard approximation, and threshold based Horvitz--Thompson variants are available when sampling fractions are non-negligible~\cite{kdd_paper_2026}. 

\subsection{Weighted Reservoir Sampling}

To efficiently draw a PPSWOR sample at scale, we use \emph{weighted random
sampling with a reservoir}~\cite{efraimidis2006weighted}. Each item $i$ is assigned an index
\begin{equation}
U_i^{1/w_{i,k}(S)},
\label{eq:reservior_sampling}
\end{equation}
where $U_i \sim \mathrm{Uniform}(0,1)$ and $w_{i,k}(S)$ is the sampling weight. The $n$ items with the largest index are selected into the sample. This procedure implements PPSWOR 
in a single streaming pass, without the need to know $N$ in advance.

\subsection{LLM Labeling}
\label{sec:llm_labeling}

A central design choice in our prevalence pipeline is the source
of ``ground truth'' labels for whether an item belongs to a
category $k \in \mathcal{K}$. Beyond the widely used LLM judges
discussed in Section~\ref{sec:related_work}, an alternative is
human expert annotation (specialists or trained raters), which
provides very high quality labels but does not scale: obtaining
tens of thousands of fresh labels per category would be much more costly~\cite{ratner2020snorkel}.

For each sampled item $i$, the LLM produces a binary
label $Z_{i,k} \in \{0,1\}$,
where $Z_{i,k} = 1$ indicates that the item is judged to belong to
category $k$, and $Z_{i,k} = 0$ otherwise.
Substituting these labels into the Hansen--Hurwitz estimator and the prevalence formula
\eqref{eq:prev_estimator_segment} yields an unbiased estimate of category prevalence as defined by the labeling rubric.
Throughout, we treat these expert validated LLM labels as the reference ground truth for $Z_{i,k}$: labeler decision quality (recall and false positive rate) is monitored and gold set gated within the reference pipeline itself~\cite{kdd_paper_2026}, and any residual label error is shared by the reference and by the surrogate calibrated against it, so it does not affect the surrogate versus reference comparisons of Section~\ref{sec:validation}. 

We use this LLM-based estimator both as
the offline calibration target for the ML score surrogate (Section \ref{sec:offline_calibration})
and as an ongoing reference against which the deployed surrogate is
audited on live experiments (Section \ref{sec:validation}).

\section{Methodology: ML Score Bucket Surrogate Prevalence}
\label{sec:methodology}

Estimating per experiment, per arm prevalence is, in principle,
a well studied problem: given a sample of labeled items, the
Hansen--Hurwitz estimator
produces an unbiased prevalence estimate, and using model scores
as auxiliary information to reduce variance is a standard
technique~\cite{holt1979post,sarndal1992model}. The challenge
addressed by our system is therefore not whether prevalence can
be estimated for any one experiment, but how to produce
\emph{thousands} of such estimates per day across hundreds of
concurrent experiments and segments, with calibration uncertainty correctly propagated, at interactive query latency, and without
incurring per experiment LLM labeling cost. This is a system level problem rather than an estimator level one. We address this by calibrating model-score buckets once and reusing
that calibration across downstream experiment cohorts.

\subsection{Bucket Level Prevalence}
\label{sec:ScoreBucketsandCalibration}
Let $m_{i,k} \in [0,1]$ be the model score for category $k$ on item $i$; higher values indicate that item $i$ is more likely to be true in category $k$. We discretize the model scores into $B$ buckets:
$$
  \mathcal{B} = \{b_1,\dots,b_B\},
$$
where each bucket $b_j$ corresponds to an interval
$$
  b_1 = [u_{0}, u_{1}),\;
  b_2 = [u_1, u_2),\;
  \dots,\;
  b_{B} = [u_{B-1}, u_{B}],
$$
with $0 = u_0 < u_1 < \dots < u_{B-1} < u_B = 1$.

For a given segment $S$ and category $k$, we define the \emph{bucket level prevalence} as
\begin{equation}
\label{eq:general_bucket_level_prevalence}
\mathcal{P}_{k,b}(S)
    \;=\;
    \mathbb{P}\!\left(Z_{k} = 1
      \,\middle|\,
      m_{k} \in b,\ \text{impression} \in S
    \right),
\end{equation}
i.e., the probability that impressions from segment $S$ are truly in
category $k$, conditional on model score $m_{k}$ falling in bucket
$b$.

\subsection{Prevalence Estimation and Variance Propagation}

Let $c_{k,b}(S)$ denote the share of impressions
from $S$ whose model scores for category $k$ fall into bucket $b$:
\begin{equation}
\label{eq:c_k_S}
\begin{aligned}
  c_{k,b}(S)
    &=         \mathbb{P}\!\left(m_{k} \in b\
      \,\middle| \ \text{impression} \in S
    \right) \\
    &=   \frac{\text{Impressions in } S \text{ with } m_{k} \in b}
         {\text{Total Impressions in } S},
\end{aligned}
\end{equation}
so that $\sum_{b \in \mathcal{B}} c_{k,b}(S) = 1$.

Using the law of total probability, we estimate the prevalence of
category $k$ in segment $S$ as:

\begin{equation}
\label{eq:prev_k_S_true}
  \widehat{\mathcal{P}}_k(S)
    \;=\;
    \sum_{b \in \mathcal{B}}
     c_{k,b}(S) \cdot \widehat{\mathcal{P}}_{k,b}(S) .
\end{equation}
In practice, we compute the calibration once on a large
``global'' segment (i.e., all traffic) and reuse the resulting
bucket level estimates $\widehat{\mathcal{P}}_{k,b}$ across all
segments $S$, yielding the segment level approximation:
\begin{equation}
  \widehat{\mathcal{P}}_k(S) \;\approx\;
    \sum_{b \in \mathcal{B}} c_{k,b}(S) \cdot
      \widehat{\mathcal{P}}_{k,b}.
  \label{eq:approx_prev}
\end{equation}
We defer the rationale for this design choice and a comparison
against the segment specific alternative to
Section~\ref{sec:global_calibration}. Intuitively, \eqref{eq:approx_prev} is a bucketed approximation
to Equation~\eqref{eq:prevalence_def}: we treat the bucket level
prevalences as a learned ``score $\to$ likelihood'' mapping, and
then reweight by the score bucket distribution of impressions in
segment $S$. 

Also, we can assume the bucket level estimators are independent across $b$ (reasonable given disjoint buckets), then applying standard variance propagation to
\eqref{eq:approx_prev} yields:
\begin{equation}
\label{eq:var_prev_k_S}
  \mathrm{Var}\!\left(\widehat{\mathcal{P}}_k(S)\right)
    \;\approx\;
    \sum_{b \in \mathcal{B}} c_{k,b}(S)^2 \,
      \mathrm{Var}\!\left(\widehat{\mathcal{P}}_{k,b}\right).
\end{equation}



\section{Production Audits Against the LLM Reference}
\label{sec:validation}

Before describing the implementation
(Section~\ref{sec:implementation}), we first report production
audits that establish the surrogate's agreement with the
LLM-based reference. Because the deployed surrogate produces a metric that product
teams act on, we run periodic audits against the LLM-based
reference estimator on live A/B
experiments, and report the results back to metric owners. Three
such audits are summarized below as case studies: a large content
filtering intervention, a UI only refresh with no expected
prevalence shift, and a subcategory targeting experiment.
Together they assess how well the surrogate recovers both
(i) the absolute prevalence levels in each arm, and
(ii) the between-group deltas.

\subsection{Experiment A: Target Category Filtering}
\label{sec:exp_A}

Our first case study covers a production A/B experiment that applies
additional filtering for two target categories, $k_1$ and $k_2$, on a major
recommendation surface. The product team's goal is to evaluate how
aggressive filtering of these categories affects user experience and
engagement; the surrogate provides their primary read on category
exposure throughout the experiment window.
The experiment includes three arms:

\begin{itemize}
  \item \textbf{Control:} baseline production settings.
  \item \textbf{Treatment$_1$:} baseline settings plus additional
        filtering for category $k_1$, removing items with
        $m_{i,k_1} \geq 0.70$ 
  \item \textbf{Treatment$_2$:} baseline settings plus additional
        filtering for category $k_2$, removing items with
        $m_{i,k_2} \geq 0.10$.
\end{itemize}

For each arm, we estimate the prevalences by:
\begin{enumerate}
  \item \textbf{LLM-based estimator (reference):}
    impression weighted sampling, and LLM labeling for each arm.
  \item \textbf{ML score surrogate estimator:}
    using the calibrated bucket prevalences $\widehat{\mathcal{P}}_{k_\ell,b}$
    for each category $k_\ell$ and bucket $b$, together with the
    arm specific impression shares $c_{k_\ell,b}(\text{Control})$ and
    $c_{k_\ell,b}(\text{Treatment}_\ell)$, aggregated according to Equation~\eqref{eq:approx_prev}. In this experiment we use $B=10$
    equally spaced buckets.
\end{enumerate}

\begin{figure*}[htbp]
  \centering
  \begin{subfigure}[b]{0.48\textwidth}
    \centering
    \includegraphics[width=\linewidth]{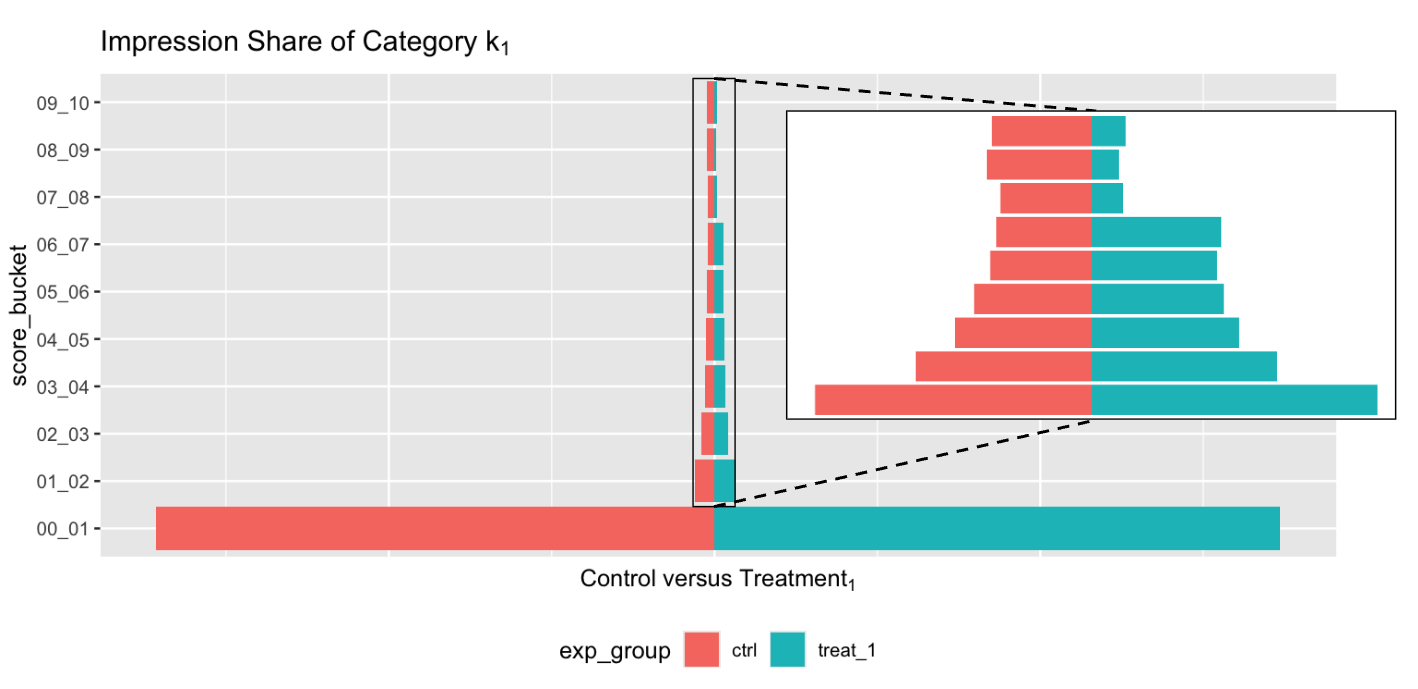}
  \end{subfigure}
  \hfill
  \begin{subfigure}[b]{0.48\textwidth}
    \centering
    \includegraphics[width=\linewidth]{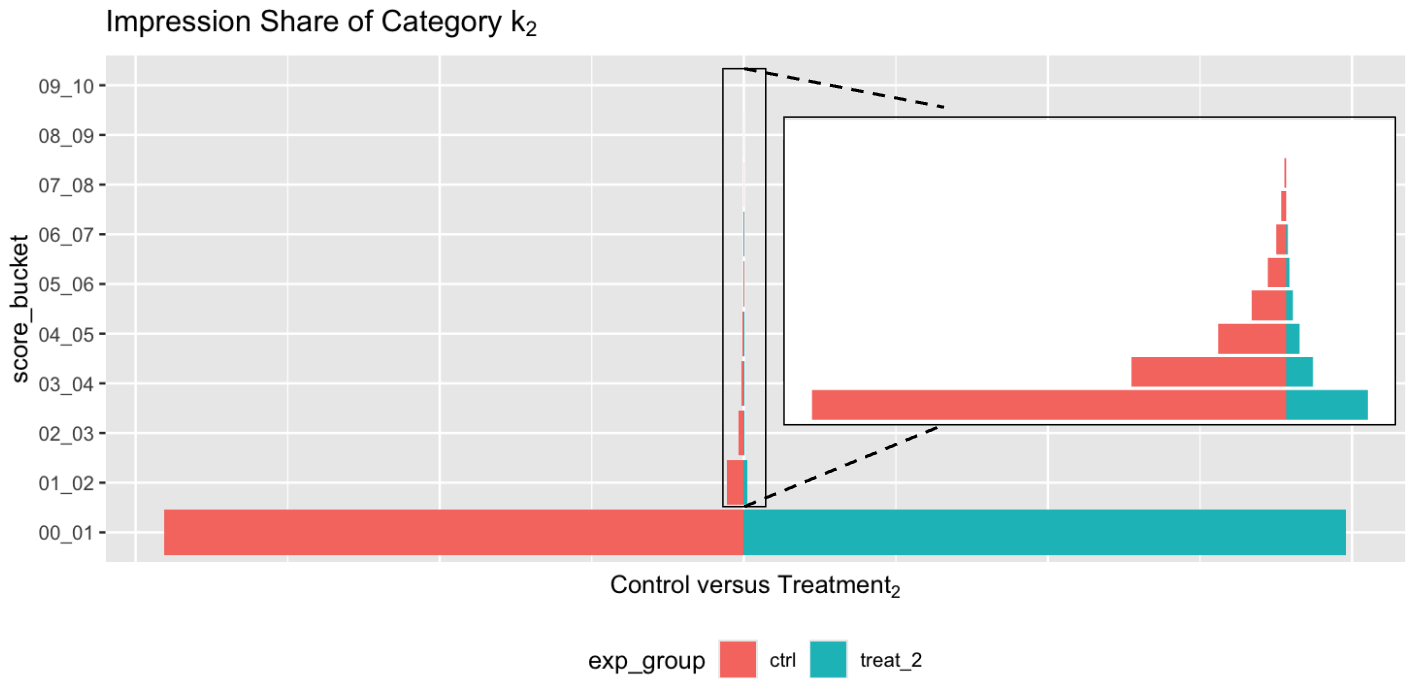}
  \end{subfigure}
  \caption{Bucket level impression share shifts for categories $k_1$ and $k_2$
           in Experiment~A. The treatment arm reduces exposure primarily by
           shifting impressions out of higher score buckets.}
  \label{fig:expA_impression_shares}
\end{figure*}

Figure~\ref{fig:expA_impression_shares} shows the impression share
distributions $c_{k_\ell,b}(\text{Control})$ (salmon) and
$c_{k_\ell,b}(\text{Treatment}_\ell)$ (cyan) over score buckets across
the entire experiment window for both categories, under the
given thresholds, respectively. Both treatments shift
impression mass away from higher-score buckets targeted by the
filtering thresholds.

Table~\ref{tab:expA_comparison} summarizes the prevalence
estimates $\widehat{\mathcal{P}}_{k_\ell}(\text{Control})$ and
$\widehat{\mathcal{P}}_{k_\ell}(\text{Treatment}_\ell)$. 
\begin{itemize}
    \item For $k_1$, the treatment reduces prevalence from
approximately 3.6\% to 2.9\% (around 19.4\% reduction) under the LLM-based estimator and from
3.8\% to 2.8\% (27.0\% reduction) under the ML score surrogate, corresponding to
treatment-control deltas of 0.70\% and 1.03\%,
respectively.
\item For $k_2$, the treatment reduces prevalence from about
1.3\% to 1.0\% (LLM; 23.5\% drop) and from 1.4\% to 0.9\% (Surrogate; 36.0\% drop), with deltas of
0.31\% and 0.49\%. 
\end{itemize}
In all cases, the surrogate's arm level estimates lie within the
95\% confidence intervals of the LLM-based reference, and the inferred
treatment effects agree in sign and magnitude.

\begin{table}[htbp]
  \centering
  \caption{Comparison of LLM-based vs.\ ML score surrogate prevalence
           in Experiment~A for categories $k_1$ and $k_2$. The ranges
           denote 95\% confidence intervals.}
  \label{tab:expA_comparison}
  \begin{tabular}{lcc}
    \toprule
    Metric & LLM-based & ML score surrogate \\
    \midrule
    $\widehat{\mathcal{P}}_{k_1}(\text{Control})$
      & 3.61\% [3.40\%, 3.82\%]
      & 3.81\% [3.67\%, 3.95\%] \\
    $\widehat{\mathcal{P}}_{k_1}(\text{Treatment}_1)$
      & 2.91\% [2.72\%, 3.10\%]
      & 2.78\% [2.64\%, 2.92\%] \\
    $\Delta_{k_1,\ \text{Treat}_1 - \text{Ctrl}}$
      & $-0.70\%$ 
      & $-1.03\%$\\
        $p$-value
      & $<0.001$ (\textbf{Stats-Sig})
      & $<0.001$ (\textbf{Stats-Sig}) \\
    \midrule
    $\widehat{\mathcal{P}}_{k_2}(\text{Control})$
      & 1.32\% [1.20\%, 1.43\%]
      & 1.36\% [1.20\%, 1.52\%] \\
    $\widehat{\mathcal{P}}_{k_2}(\text{Treatment}_2)$
      & 1.01\% [0.91\%, 1.11\%]
      & 0.87\% [0.70\%, 1.04\%] \\
    $\Delta_{k_2,\ \text{Treat}_2 - \text{Ctrl}}$
      & $-0.31\%$
      & $-0.49\%$\\
    $p$-value
      & $<0.001$ (\textbf{Stats-Sig})
      & $<0.001$ (\textbf{Stats-Sig}) \\
    \bottomrule
  \end{tabular}
\end{table}

\subsection{Experiment B: UI Only Change with No Expected Prevalence Shift}
\label{sec:exp_B}

Our second case study covers a user interface refresh where we do not
expect any change in category $k_1$ and $k_2$ prevalence. 
The experiment updates the web overflow menu by modifying the hide/report icon, without altering the underlying
functionality or ranking logic.


Table~\ref{tab:expB_comparison} summarizes the estimates over the whole experiment window. For $k_1$, the control and treatment arms have very
similar prevalences: 3.44\% vs.\ 3.31\% under the LLM-based estimator
and 3.39\% vs.\ 3.40\% under the ML score surrogate. For $k_2$, the
prevalences are likewise close: 1.32\% vs.\ 1.40\% (LLM) and 1.39\% vs.\
1.35\% (surrogate), with treatment-control deltas near zero in all cases.

The surrogate estimates lie within the 95\% confidence intervals of the LLM-based reference, and none of the inferred treatment effects are statistically significant.

\begin{table}[htbp]
  \centering
  \caption{Comparison of LLM-based vs.\ ML score surrogate prevalence for categories $k_1$ and $k_2$ in Experiment~B.}
  \label{tab:expB_comparison}
  \begin{tabular}{lcc}
    \toprule
    Metric & LLM-based& ML score surrogate \\
    \midrule
    $\widehat{\mathcal{P}}_{k_1}(\text{Control})$
      & 3.44\% [3.23\%, 3.64\%]
      & 3.39\% [3.25\%, 3.53\%]\\
    $\widehat{\mathcal{P}}_{k_1}(\text{Treatment})$
      & 3.31\% [3.11\%, 3.51\%]
      &  3.40\% [3.26\%, 3.53\%]\\
    $\Delta_{k_1,\ \text{Treat} - \text{Ctrl}}$
      & $-0.13\%$
      & $0.01\%$\\
     $p$-value
      & $0.38$ (\textbf{No Stats-Sig})
      & $0.95$ (\textbf{No Stats-Sig}) \\   
    \midrule
    $\widehat{\mathcal{P}}_{k_2}(\text{Control})$
      & 1.32\% [1.19\%, 1.45\%]
      & 1.39\% [1.22\%, 1.56\%] \\
    $\widehat{\mathcal{P}}_{k_2}(\text{Treatment})$
      & 1.40\% [1.26\%, 1.53\%]
      & 1.35\% [1.18\%, 1.52\%] \\
    $\Delta_{k_2,\ \text{Treat} - \text{Ctrl}}$
      & $+0.08\%$
      &   $-0.04\%$ \\
    $p$-value
      & $0.40$ (\textbf{No Stats-Sig})
      & $0.67$ (\textbf{No Stats-Sig}) \\   
    \bottomrule
  \end{tabular}
\end{table}

\subsection{Experiment C: Detecting a Small Production Shift Beyond the Reach of a Single LLM Measurement} 
\label{sec:exp_C}

Our third case study covers an experiment that
targets a subcategory of category $k_1$. Because this subcategory accounts for only a small fraction of the overall $k_1$ impression share, any change in $\mathcal{P}_{k_1}$ is expected to be modest but systematic. This is the regime in which the deployed surrogate is most useful:
the per arm calibration uncertainty of any single LLM measurement is wide compared with the expected effect, while the surrogate can be re-evaluated daily without any additional label cost.

Aggregated throughout the experiment window, the LLM-based
estimator yields $
  \widehat{\mathcal{P}}_{k_1}(\mathrm{control})
    \approx 3.83\%$, and $\widehat{\mathcal{P}}_{k_1}(\mathrm{treatment})
    \approx 3.67\%$
with a two sided $p\text{-value} \approx 0.31$ and overlapping 95\% confidence intervals $[3.69\%, 3.96\%]$ and $[3.53\%, 3.80\%]$, respectively.

\begin{figure}[t]
  \centering
  \includegraphics[width=0.48\textwidth]{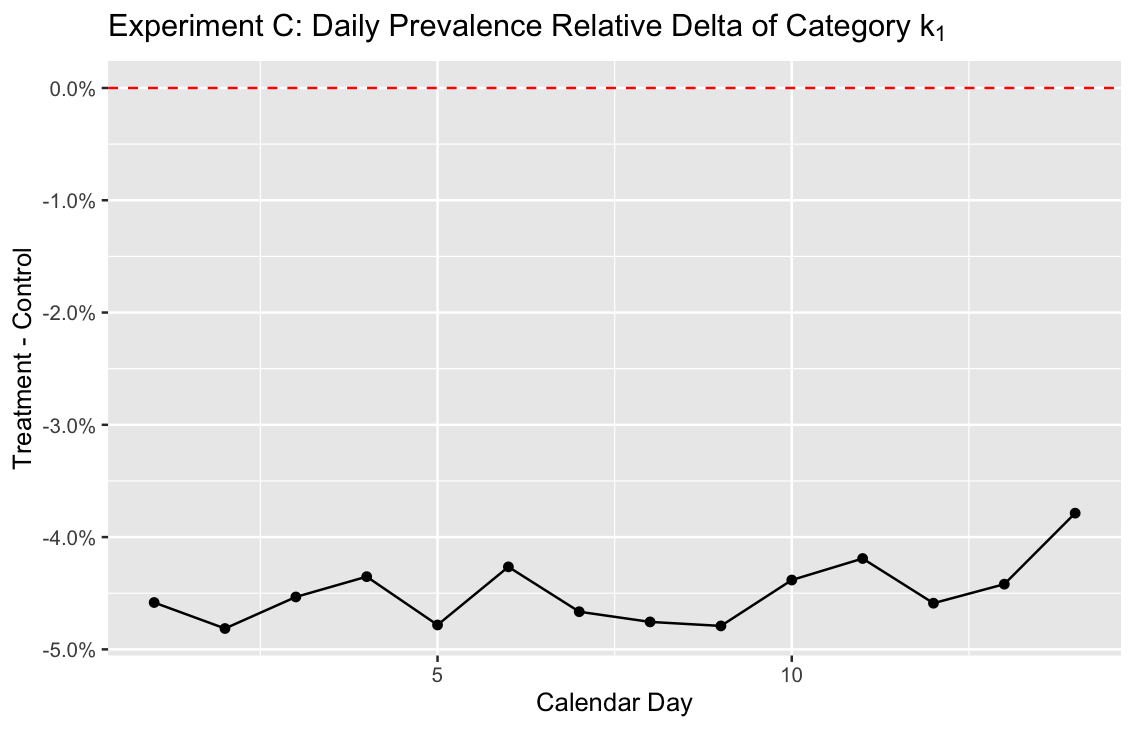}
  \caption{Relative prevalence reduction vs. calendar day: $\Delta(d)/ \widehat{\mathcal{P}}_{k_1}(\mathrm{control};d)$
for category $k_1$ in Experiment~C.}
  \label{fig:expc_daily_delta}
\end{figure}

The deployed surrogate, in contrast, computes a daily prevalence
per arm and the corresponding daily effect \begin{equation}
      \Delta(d)
    = \widehat{\mathcal{P}}_{k}(\mathrm{treatment}; d)
      - \widehat{\mathcal{P}}_{k}(\mathrm{control}; d),
      \label{eq:delta}
\end{equation} 
under a fixed calibration snapshot. Aggregating these day level deltas using a sign test~\cite{dixon1946statistical} (Section \ref{sec:implementation_stage_2}) gives a mean delta of approximately $-0.16\%$, a $4.2\%$ relative reduction on a baseline (control) prevalence of
about $3.8\%$ (using ML surrogate), with daily values that are consistently negative
across the experiment window (Figure \ref{fig:expc_daily_delta}). The corresponding day level
sign-test $p\text{-value} < 0.001$, indicating a statistically
significant shift at the experiment level.

As a sanity check and to guard against false positives, we also apply the same day level aggregation to Experiment~B, where no prevalence
shift is expected. This yields $p\text{-value}_{k_1} \approx 0.75$ and $p\text{-value}_{k_2} \approx 0.25$, aligned with the null hypothesis.

This case study illustrates a production relevant sensitivity gap: a small but systematic around $-4\%$ relative reduction is invisible to a single per arm LLM measurement at the labeling budget we can afford, but is recovered by the deployed surrogate via day level aggregation.

\section{Implementation}
\label{sec:implementation}

We now describe how the surrogate is deployed in production: an
offline calibration pipeline that periodically refreshes
$\{\widehat{\mathcal{P}}_{k,b}\}$ from the always-on LLM
pipeline's label stream
(Section~\ref{sec:offline_calibration}), a daily online
integration with the experiment platform that materializes
per experiment prevalences and renders the dashboard
(Section~\ref{sec:online_intergration}), and the operational
cadence under which both run
(Section~\ref{sec:operations}).

\subsection{Offline Calibration Pipeline}
\label{sec:offline_calibration}

The calibration step estimates the bucket-level prevalences
$\{\widehat{\mathcal{P}}_{k,b}\}_{b \in \mathcal{B}}$ and their variances
$\{\mathrm{Var}(\widehat{\mathcal{P}}_{k,b})\}_{b \in \mathcal{B}}$ that the online metric consumes. In Pinterest's deployment, these
quantities are computed from the labeled stream already produced by
the daily batch LLM prevalence workflow, which maintains a global
PPSWOR sample via weighted reservoir sampling
$w_{i,k} \propto f(I_{i})$. Reusing this stream gives the surrogate
zero incremental labeling cost.

Given the sampled and LLM-labeled stream, we estimate each
bucket-level prevalence by applying the Hansen--Hurwitz estimator
with the bucket itself as the segment, $S=b$:
\begin{equation}
  \widehat{\mathcal{P}}_{k,b}
  \;=\; \widehat{\mathcal{P}}_{k}(S = b).
  \label{eq:bucket_prevalence_def}
\end{equation}
Concretely, let $I_i(b)$ be the number of impressions of item $i$ arising from bucket $b$ in the calibration window, and let $\mathbb{1}_{i \in b}$ be the
indicator that item $i$ receives at least one impression from $b$. We estimate the joint probability that an impression is both in category $k$ and in bucket $b$ by normalizing the Hansen--Hurwitz total by platform impressions:
\begin{equation}
  \widehat{P}\!\bigl(Z_{i,k} = 1 \cap
                     \text{impression} \in b\bigr)
  \;=\; \frac{1}{\sum_{i \in \mathcal{D}} I_i}\cdot\frac{1}{n}\sum_{t=1}^{n}
          \frac{Z_{i_t,k}\,
                \mathbb{1}_{i_t \in b}\,
                I_{i_t}(b)}
               {p_{\mathrm{sample},i_t,k}}.
  \label{eq:joint_prob}
\end{equation}

Then, the marginal probability
that a random impression falls in bucket $b$ is computed
directly from logs:
          $P(\text{impression} \in b)
          \;=\; \frac{\sum_{i \in \mathcal{D}(b)} I_i(b)}
                     {\sum_{i \in \mathcal{D}} I_i}$.
By the definition of conditional probability, the estimated prevalence of category $k$ in bucket $b$ is then
\begin{equation}
  \widehat{\mathcal{P}}_{k}(S = b)
  \;=\; \frac{\widehat{P}\!\bigl(Z_{i,k} = 1 \cap
                                  \text{impression} \in b\bigr)}
             {P(\text{impression} \in b)}.
  \label{eq:bucket_prevalence}
\end{equation}

\begin{figure*}[htbp]
  \centering
  \begin{subfigure}[b]{0.48\textwidth}
    \centering
    \includegraphics[width=\linewidth]{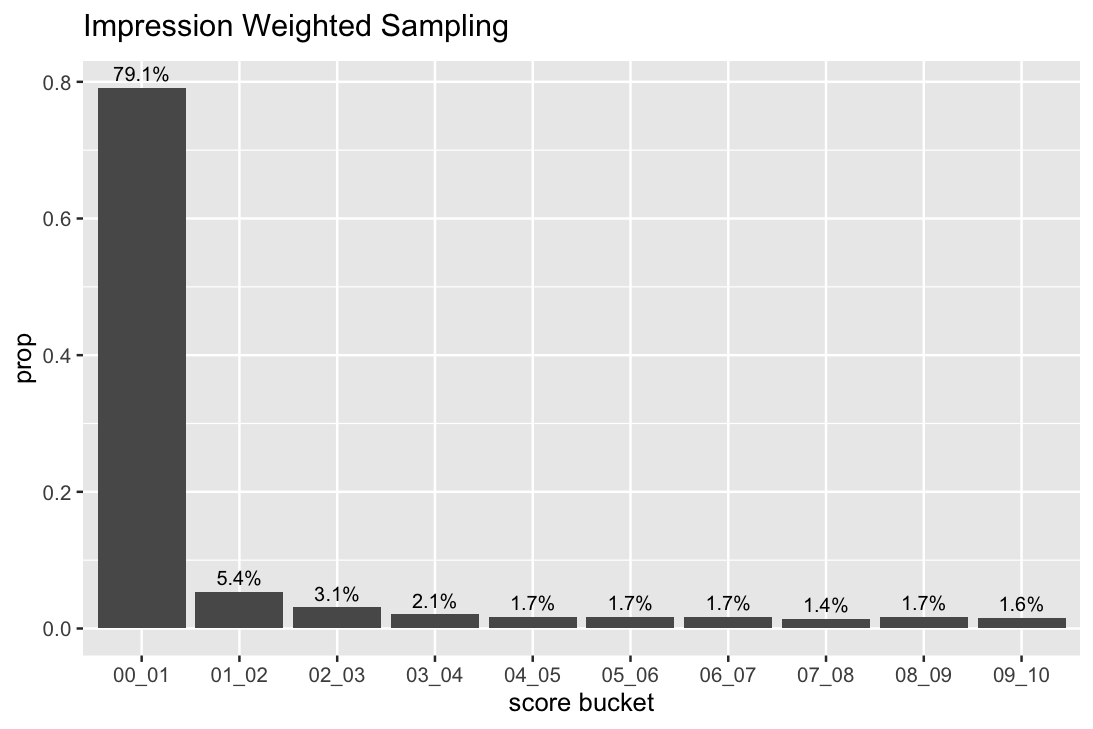}
    \caption{Impression weighted sampling ($w_{i,k} \propto I_i$).}
  \end{subfigure}
  \hfill
  \begin{subfigure}[b]{0.48\textwidth}
    \centering
    \includegraphics[width=\linewidth]{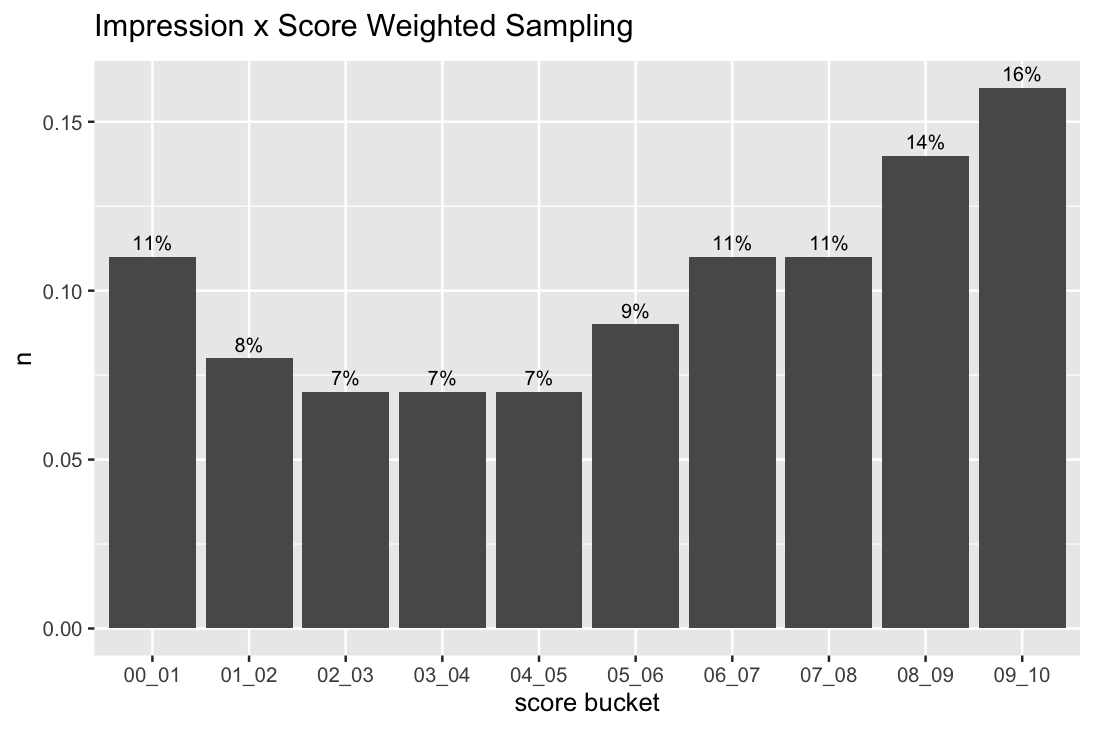}
    \caption{Impression $\times$ score sampling ($w_{i,k} \propto I_i m_{i,k}$).}
  \end{subfigure}
  \caption{Comparison of score bucket distributions under two sampling schemes. Left: sampling with weights proportional to
           impressions alone produces a very low score heavy sample,
           leaving few examples in high score buckets. Right: sampling
           with weights proportional to impression $\times$ model score
           yields a more balanced distribution across buckets, improving
           the precision of bucket level prevalence estimates.}
  \label{fig:distribution_impression_score_sampling}
\end{figure*}

A practical concern in this shared sampling design is that model
scores are highly skewed toward low values. If the always-on
pipeline used impressions alone as weights, i.e.,
$w_{i,k} \propto I_i$, the resulting score distribution in the
sample would be dominated by the lowest bucket: in our
data, as shown in Figure~\ref{fig:distribution_impression_score_sampling} (a), roughly $80\%$
of sampled impressions fall into the bucket $[0, 0.1)$, while buckets with
larger scores (e.g., $m_{i,k} > 0.3$) each contain only about
$2\%$ of the mass. With a sample of $10{,}000$ items, fewer than
$200$ would lie in each higher bucket, limiting statistical power
for estimating $\mathcal{P}_{k,b}$ at the upper end of the score
range.

The always-on pipeline therefore uses the model score as auxiliary
information~\cite{sarndal1992model}, i.e.,
$w_{i,k} \propto I_i\, m_{i,k}$, which boosts the inclusion
probability of higher score items. On the same underlying dataset, shown in Figure~\ref{fig:distribution_impression_score_sampling} (b), this design yields a much more
balanced bucket distribution: every bucket has at least
$\sim\! 7\%$ of the sample, and the two highest score buckets account
for roughly $14\%$ and $16\%$. Although this weighting was originally chosen
so that the LLM-based reference estimator has good precision in the
high score tail, the surrogate inherits a sample design that is
already well suited to bucket level calibration, with minimal
operational overhead.

\subsection{Online Integration}
\label{sec:online_intergration}

The online side of the system runs as a two stage pipeline
integrated with Pinterest's experimentation platform: a daily
per experiment prevalence computation job, followed by a dashboard
layer that produces the experiment level read from the day level
results.

\subsubsection{Stage 1: daily prevalence per experiment} 
\label{sec:implementation_stage_1}
For each qualified
experiment (currently those run by the Content Quality,
Trust \& Safety, and Overall Holdout programs), a scheduled
job runs once per day per arm and:
\begin{enumerate}
  \item joins the experiment assignment table (which users are in
      which arm) with the daily impression logs to obtain
      per (arm, day) impressions,
  \item pulls model scores for the calibrated category $k$ on those impressions,
  \item aggregates to obtain the impression shares
      $c_{k,b}(S; d)$ per bucket per arm per day,
  \item reads the latest calibration snapshot $\{\widehat{\mathcal{P}}_{k,b}\}_{b \in \mathcal{B}}$ and computes the arm level prevalence $\widehat{\mathcal{P}}_{k}(S; d)$ together with its analytic 95\% confidence interval.
\end{enumerate}

The impression and assignment join is resolved here, in the
scheduled job, rather than at query time. Each (experiment, arm,
day) row is therefore computed once and reused by every
downstream consumer.

\subsubsection{Stage 2: dashboard rendering}\label{sec:implementation_stage_2} The experimentation platform dashboard consumes the prevalence
table from \emph{Stage~1} and produces the experiment level verdict
using a delta focused inference layer. 

The motivation is that
the per arm analytic confidence intervals are often dominated by the
bucket level calibration variance
$\mathrm{Var}(\widehat{\mathcal{P}}_{k,b})$ rather than by the
experiment level variation in the bucket mixes $c_{k,b}(S)$.
With a sufficiently large calibration sample, this calibration variance could be
made negligible, but in practice our calibration budget is
constrained by LLM labeling cost. 
At realistic calibration sizes
the per arm CIs for $\widehat{\mathcal{P}}_k(\text{control})$ and
$\widehat{\mathcal{P}}_k(\text{treatment})$ are therefore best
interpreted as answering: \emph{``If we repeated the global LLM
calibration many times, where would the absolute prevalence for
this arm lie?''}; a useful question when the experimental shift
is large (e.g., Experiment~A in
Section~\ref{sec:exp_A}, where prevalence moves by roughly 20\%
or more), but a poor question in the more common production
setting where a 2\%--5\% relative change is already meaningful (e.g., Experiment~C in
Section~\ref{sec:exp_C}).

In that regime, calibration uncertainty can dominate the per arm
CI and cause small but systematic treatment effects to appear
statistically non-significant.
The dashboard therefore reports inference computed on the
day level deltas $\Delta(d)$ (Equation \ref{eq:delta}). 
Concretely, on day $D$ the dashboard collects the
trailing window of daily deltas
$\{\Delta(D-m), \ldots, \Delta(D-1)\}$ for the experiment arm
pair (we use minimum $m = 10$ days in production) and reports:

\begin{itemize}
  \item the mean delta $\bar{\Delta}$ over the window,
  \item an empirical 95\% confidence interval
        $[\Delta_{0.025}, \Delta_{0.975}]$ from the quantiles
        of $\{\Delta(d)\}$, and
  \item a sign-test $p$-value~\cite{dixon1946statistical} based on the
        fraction of days on which $\Delta(d)$ is positive versus
        negative.
\end{itemize}

If the sign-test $p$-value is below the chosen significance
level the observed mean effect $\bar{\Delta}$ is unlikely to be
explained by random variation under the null hypothesis of no
daily effect (Figure \ref{fig:expc_daily_delta}); otherwise, we fail to reject the null. These
statistics are surfaced to the experiment owner as the headline
prevalence read for day $D$. For experiments
shorter than the $m$-day minimum, only the day level prevalences
and a per day delta are shown. Because \emph{Stage~1} has already
materialized the prevalence rows the dashboard depends on, this
dashboard query is effectively a small read over precomputed
numbers and returns at interactive latency ($<$10 seconds; Table~\ref{tab:operation}).

\subsection{Operations}
\label{sec:operations}

The cost structure depends on whether a platform already operates a
recurring LLM prevalence workflow. In Pinterest's deployment, the
daily batch LLM labeling pipeline is already funded as part of the
platform-level reference measurement system, so the surrogate adds
only calibration computation and the per-experiment SQL metric, with
zero incremental labeling cost. For teams without such a workflow,
the same architecture requires a recurring global calibration-labeling
job; the labeling cost is not zero, but it is paid once at the platform
or category level and amortized across all enrolled experiments.

The relevant comparison is per-experiment LLM labeling. At roughly
$60{,}000$ samples per arm per day, labeling every cell would require on the order of
$10^9$ labels per category per year, which is operationally infeasible.
In practice, the realistic alternative is a one-shot LLM measurement
per arm for only a small fraction of experiments under the available
labeling budget (Table \ref{tab:operation}).
\begin{table}[htbp]
  \centering
  \caption{Relative Operational Comparison: LLM vs. Surrogate}
  \label{tab:operation}
  \begin{tabular}{lcc}
    \toprule
    Metric & LLM-based& ML score surrogate \\
    \midrule
    \# of Measured Exp
      & Baseline
      & $>20\times$\\
    Extra Labeling Cost
      & Baseline
      & \$0\\
    Frequency
      & One Run Per Exp
      & Daily Run Per Arm\\
    Latency
      & 10 to 24 Hours
      & <10s\\
    \bottomrule
  \end{tabular}
\end{table}

Even on those
covered arms, a single LLM measurement carries its own sampling uncertainty: small but systematic
shifts in the $2$--$5\%$ relative range, the regime that
dominates the production caseload, are typically statistically
indistinguishable from noise within a one shot confidence interval.
The deployed surrogate closes both gaps: it provides per arm
prevalence on every enrolled arm, and by re-evaluating daily under a fixed calibration
snapshot, it accumulates day level deltas that recovers
small effects that any single LLM measurement would miss.

Audits against the LLM-based reference are run on two cadences. 
\begin{itemize}
    \item Daily platform level prevalence is audited every day, since the
always-on LLM pipeline produces a daily platform measurement that
the surrogate can be compared against without additional labeling
cost. 
\item Per arm audits on enrolled A/B experiments are run
periodically, at program onboarding and on a recurring
cadence, since each per arm audit requires fresh LLM labeling of
arm specific samples. 
\end{itemize}
Across approximately 300 such audits in total, the surrogate's 95\% confidence interval
contains the LLM-based reference point estimate in $92\%$ of
evaluations, and the surrogate's CI overlaps the LLM-based
reference's CI in $96\%$ of evaluations.
Note that the LLM-based reference point estimate itself carries
sampling uncertainty, so even a perfectly calibrated surrogate is
expected to contain the reference \emph{point} less often than the
nominal $95\%$ level; the $96\%$ CI overlap rate is the more directly
interpretable calibration figure.

\section{Discussion and Future Work}
\label{sec:discussion}
\subsection{Choice of Score Buckets}
\label{sec:bucket_choice}
The deployed system supports both fixed and adaptive score
bucketization. For some calibrated categories, we use a simple default
of $B=10$ equal-width buckets over $[0,1]$, combined with the
impression$\times$score weighted sampling described in
Section~\ref{sec:offline_calibration}. This default is easy to
implement and stable across daily refreshes, but it can be unreliable
for skewed score distributions and rare positives: many samples may
pile up near zero while higher-score buckets receive too few labels or
too few positives to support stable bucket-level prevalence estimates.

For such categories, we use adaptive binning in production. The method
starts from equal-frequency score micro-bins, then merges neighboring
micro-bins until each retained bucket satisfies minimum sample and
positive-label thresholds. If too many buckets remain, it merges the
adjacent pair with the smallest information loss, favoring merges
between bins with similar empirical positive rates or small sample
sizes. The final boundaries are rounded and cleaned to cover $[0,1]$
and remain stable at scoring time.

We treat bucketization as a deployment choice rather than an
estimator-level contribution. Equal-width buckets are sufficient when
stable; adaptive buckets are used when rare positives or score skew
make fixed-width calibration unreliable. More broadly, the surrogate
can represent content or context features beyond raw model score, and
the appropriate bucketization strategy depends on the application,
calibration budget, and operational stability requirements.

\subsection{Choice of Calibration Granularity}
\label{sec:global_calibration}
\begin{figure}[htbp]
  \centering
  \includegraphics[width=\linewidth]{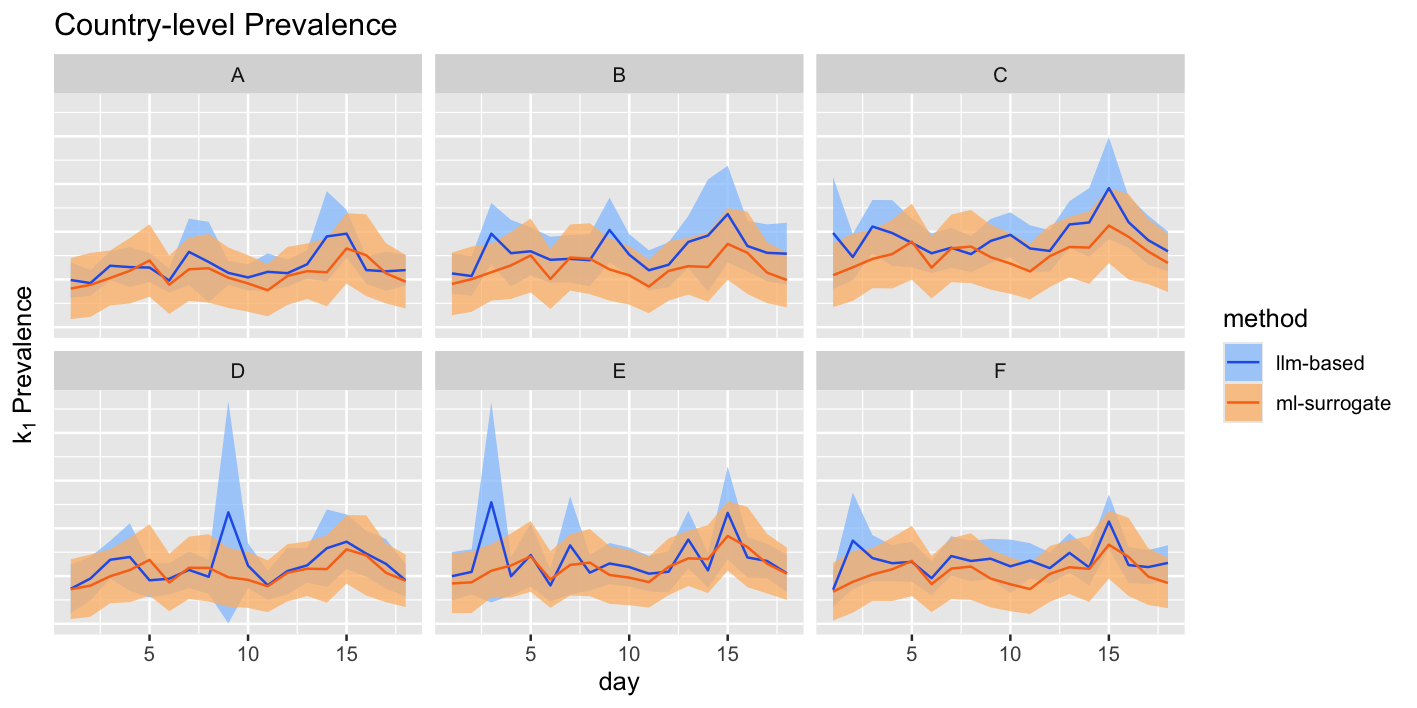}
  \caption{Daily $k_1$ prevalence for the six highest-engagement
    countries (A--F), comparing the LLM-based estimator and the
    ML-score surrogate over a 30-day window; ribbons denote 95\%
    confidence intervals.}
  \label{fig:country_prevalence}
\end{figure}

The deployed system currently applies a single global calibration
$\widehat{\mathcal{P}}_{k,b}$ to every segment $S$, even though the
framework of Section~\ref{sec:methodology} also supports
segment-specific calibrations $\widehat{\mathcal{P}}_{k,b}(S)$. This
choice works well when, for a fixed category $k$ and bucket $b$, the
bucket-level prevalence $\mathcal{P}_{k,b}(S)$ is stable across the
segment family of interest. In that regime, segment differences are
mainly captured by the impression-share distribution $c_{k,b}(S)$,
while the shared calibration pools labels
across segments and reduces noise. We observe this pattern for many reads, such as surface-level
and country-level. Figure~\ref{fig:country_prevalence}
illustrates the country case: the LLM-based country estimates and the
surrogate agree on absolute level, while the surrogate produces smoother
per-country trajectories because its country-specific component comes
from full impression logs and its label-derived component is pooled
globally.

Global calibration is not universally optimal. For some segment
families, such as top-level content verticals, the conditional
prevalence within the same score bucket can differ materially across
segments; for example, GenAI prevalence in an Art vertical may differ
from that in Home Decor at the same model-score bucket. In such cases,
a global calibration can introduce segment-specific bias, and
calibration granularity should be chosen empirically. A future extension
is a bucket-level heterogeneity check for each category $k$, segment
family $\{S_1,\ldots,S_J\}$, and bucket $b$:
\[
H_0:\mathcal{P}_{k,b}(S_1)=\cdots=\mathcal{P}_{k,b}(S_J).
\]
This could be implemented using Wald or $\chi^2$ tests over the
Hansen--Hurwitz bucket estimates, with multiple-testing correction
across buckets. If heterogeneity appears in high-mass buckets and each
segment has enough labels and positives, segment-specific calibration
may be preferable; otherwise, global calibration remains more stable.

We have not yet deployed automatic calibration-granularity selection.
A natural middle ground is partial pooling, which shrinks
segment-specific $\widehat{\mathcal{P}}_{k,b}(S)$ toward the global
$\widehat{\mathcal{P}}_{k,b}$, capturing real segment-level differences
when supported by data while avoiding high variance in sparse
segment-bucket cells.

\subsection{Outlook on LLM Cost Trends}
\label{sec:future_outlook}
A key motivation for the deployed system is that large scale LLM labeling is currently too expensive to serve as a default per experiment metric. In the longer term, as LLM inference
becomes cheaper and more tightly integrated into serving stacks,
it may become feasible to label much larger fractions of the
corpus (or even all content) for multiple content attributes.

Viewed through a tokenomics lens, the surrogate is a general
mechanism for mapping expensive LLM decisions onto cheaper models,
in the spirit of LLM cascades~\cite{chen2024frugalgpt} and
knowledge distillation~\cite{hinton2015distilling}.
The cheap signal need not be a pre existing production score: a
low cost open source model can play the same role, with its bucket
calibration periodically reinforced by a more capable, more
expensive model. In our deployment, existing ML scores provide this
signal at no additional inference cost, and the payoff is coverage:
the reference pipeline labels on the order of one million items per
day for platform level prevalence monitoring~\cite{kdd_paper_2026},
while the calibrated surrogate extends measurement to the billions
of daily impressions already recorded in logs, tightening confidence
intervals and unlocking segment level deep dives and per experiment
A/B reads that per item LLM labeling could never cover.

In that regime, the relative importance of the surrogate may
decrease for absolute prevalence estimation. However, the contribution of this work is system level: amortizing one calibration across hundreds of concurrent experiments, materializing per experiment results in a single SQL metric, and separating absolute levels from deltas in the dashboard layer. We expect these ideas to remain useful even when LLM labeling is no longer the primary bottleneck; at that point, the contents of the calibration table can change, but the system pattern carries over.

\begin{acks}
We thank the following collaborators for their support and feedback throughout this project:
Minli Zang, Benjamin Thompson, Xiaohan Yang, Wenjun Wang, Huan Yu, Faisal Farooq, Darren Reger, Andrey Gusev, Aravindh Manickavasagam,  Qinglong Zeng, Jianjin Dong, Ziming Yin,  Cindy Zhang, Gerardo Gonzalez, Ahmed Fayez, Yasmin ElBaily and Sari Wang.
\end{acks}

\bibliographystyle{ACM-Reference-Format}
\bibliography{score_bucket}

\appendix

\end{document}